\documentclass[a4paper,11pt]{article}
\usepackage{pos}

\def \arcmin {\hbox{$^\prime$}}

\def \gray {$\gamma$-ray}

\def \apj {ApJ}
\def \apjl {ApJL}

\def \aap {A\&A}

\def \nat {Nature}

\title{The ASTRI Mini-Array Core Science Program}

\manuallySeparateAuthors
\author*[a]{Stefano Vercellone}

\author[b]{ for the ASTRI Project}

\affiliation[a]{INAF -- Osservatorio Astronomico di Brera,\\
  Via E. Bianchi 46, 23807 Merate (LC), Italy}

\affiliation[b]{\href{http://www.astri.inaf.it/en/library/}{http://www.astri.inaf.it/en/library/}\\
}

\emailAdd{stefano.vercellone@inaf.it}

\abstract{Celestial sources emitting at high-energy (HE, E \textgreater 100 MeV) and at very high-energy (VHE, E \textgreater 100 GeV) gamma-rays are of the order of a few thousands and a few hundreds, respectively. On the other hand, the number of sources emitting at ultra high-energy (UHE, E \textgreater{} several tens of TeV) gamma-rays are just a few dozens, and are currently being investigated by means of both ground-based imaging atmospheric Cherenkov telescopes (IACTs) and particle shower arrays. These rare VHE and UHE sources represent a new frontier in astrophysics. An array composed of nine ASTRI Cherenkov telescopes is under construction at the Observatorio del Teide (Tenerife, Spain). The ASTRI Mini-Array aims at providing robust answers to a few selected open questions in the VHE and UHE domains. The scientific program during the first four observing years will be devoted to the following Core Science topics: the origin of cosmic rays, the extra-galactic background light and the study of fundamental physics, the novel field in the VHE domain of gamma-ray bursts and multi-messenger transients, and finally the use of the ASTRI Mini-Array to investigate ultra high-energy cosmic rays and to undertake stellar intensity interferometry studies. We review the scientific prospects assessed through dedicated simulations, proving the potential of the ASTRI Mini-Array in pursuing breakthrough discoveries and discuss the synergies with current and future VHE facilities in the Northern hemisphere, such as MAGIC, LHAASO, HAWC, Tibet AS-$\gamma$, and CTAO-N.}

\FullConference{%
  7th Heidelberg International Symposium on High-Energy Gamma-Ray Astronomy (Gamma2022)\\
  4-8 July 2022\\
  Barcelona, Spain\\}

\begin{document}
\maketitle

%
%
\section{Introduction: The ASTRI Mini-Array}
%
%
The ASTRI Mini-Array of imaging atmospheric Cherenkov telescopes~\citep{2022JHEAp..35...52S,GiulianiThisProc} is currently being installed at the Observatorio del Teide (island of Tenerife, Spain) thanks to an agreement between INAF and the {\it Instituto de Astrofisica de Canarias}. The starting activities will be managed by the {\it Fundaci\'on Galileo Galilei-INAF}. The ASTRI Mini-Array includes national and international partners.\footnote{On the Italian side, the ASTRI Collaboration encompasses the Universities of Perugia, Padova, Catania, Genova and the Milano Polytechnic together with the INFN sections of Roma Tor Vergata and Perugia. On the international side, in addition to the strategic partnership with IAC, the ASTRI Collaboration includes the University of S\~ao Paulo with FAPESP in Brasil, the North Western University in South Africa and the University of Geneva in Switzerland.}
The ASTRI Mini-Array will provide a fully functional complement for MAGIC and CTAO North. In particular, the ASTRI Mini-Array is expected to improve the MAGIC sensitivity in the North for E > few TeV and, at the same time, to operate for a few years before the full completion of CTAO North. Therefore, the ASTRI Mini-Array will have a vast discovery space in the field of extreme gamma-rays, up to 100\,TeV and beyond, as described in~\citep{2022JHEAp..35....1V,2022JHEAp..35..139D,2022JHEAp..35...91S}.
We can summarize the main ASTRI Mini-Array performance as follows: wide field of view (FoV) of about 10\textdegree{}; energy range from 1\,TeV up to 200\,TeV, angular resolution $\sim 3$\arcmin{} at 10\,TeV and  energy resolution $\sim 10$\% at 10\,TeV\footnote{We point out that, with the use of appropriate analysis cuts, angular and energy resolution can be improved throughout the entire energy range above a few TeV~\citep{2022icrc.confE.884L}. This appears to be particular relevant for morphological and spectral studies in the 1-100 TeV energy range.}.  A detailed description of the ASTRI Mini-Array performance is provided in~\citep{2022icrc.confE.884L}.
%
%
\begin{figure}[!ht]
	\vspace{-0.7cm}
	\centering
	\includegraphics[width=0.35\textwidth, angle=90]{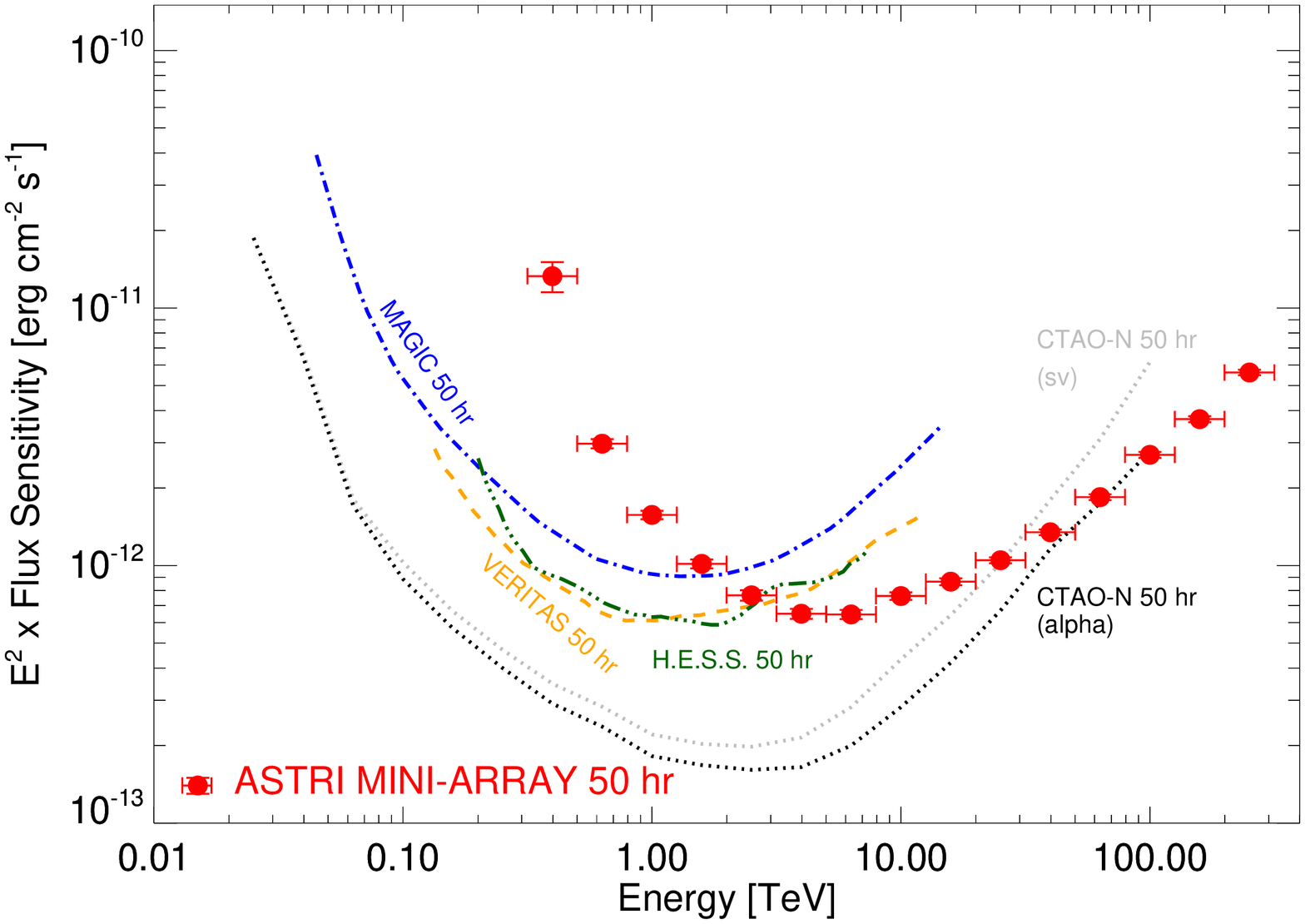}
        \includegraphics[width=0.35\textwidth, angle=90]{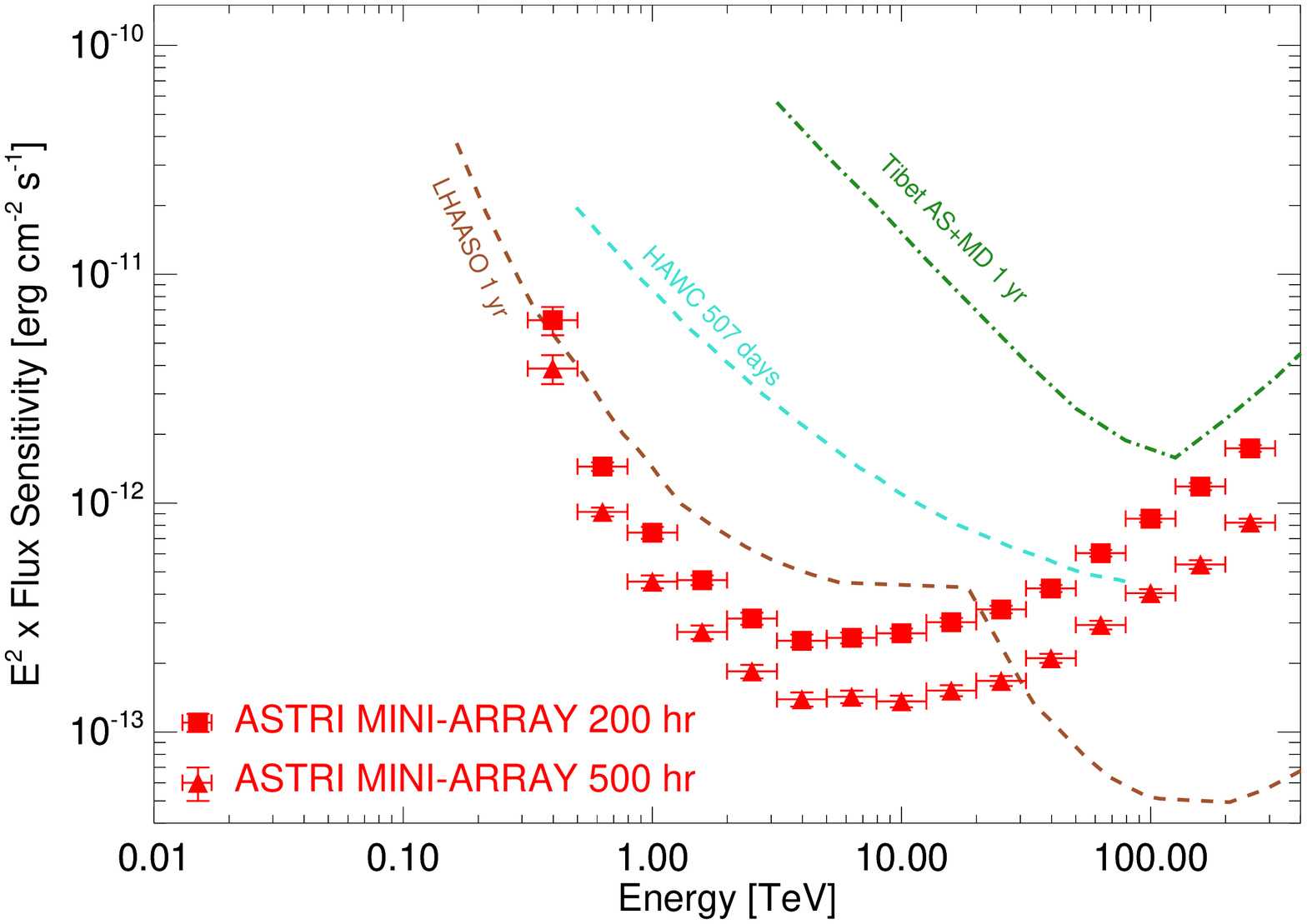}
	\caption{{\it Left Panel:} ASTRI Mini-Array differential sensitivity for 50\,hr integration compared with those of current IACTs. {\it Right Panel:} ASTRI Mini-Array 200\,hr and 500\,hr differential sensitivity compared with those of current EASs in the northern hemisphere. Both panels adapted from~\cite{2022JHEAp..35....1V}. CTAO North differential sensitivity curves come from \citep[][(alpha)]{2022icrc.confE.885G} and \citep[][(sv)]{2021_Zanin_Priv_Comm}, respectively.}
		\label{FIG:IACT_WCDA}
\end{figure}

Figure~\ref{FIG:IACT_WCDA} (left panel) shows the ASTRI Mini-Array differential sensitivity (50\,hr integration time) compared with those of current and planned imaging atmoheric Cherenkov telescope arrays (IACTs). The differential sensitivity curves come from \citep[][ASTRI Mini-Array]{2022icrc.confE.884L}, \citep[][MAGIC]{2016APh....72...76A}, the VERITAS official website\footnote{\href{https://veritas.sao.arizona.edu}{https://veritas.sao.arizona.edu}}, and \citep[][sensitivity curve for H.E.S.S.--I, stereo reconstruction]{2015ICRC...34..980H}. On the right panel, we plot the ASTRI Mini-Array differential sensitivity compared with those of current extended air-showers arrays (EASs) in the northern hemisphere. The integration times are 200\,hr and 500\,hr for the ASTRI Mini-Array (typical for a very deep pointing during the core science phase) and about 1\,yr for EASs, respectively. The differential sensitivity curves come from \citep[][ASTRI Mini-Array]{2022icrc.confE.884L}, \citep[][HAWC]{2017ApJ...843..116A}, and \citep[][LHAASO]{2016NPPP..279..166D}.

In the left panel we also plotted for comparison the CTAO North differential sensitivity curves for 50\,hr integration time both in the ``alpha''~\citep[(4\,LSTs\,$+$\,9\,MSTs),][]{2022icrc.confE.885G} and in the ``science verification''~\citep[(4\,LSTs\,$+$\,5\,MSTs),][]{2021_Zanin_Priv_Comm} configurations, respectively. The ASTRI Mini-Array differential sensitivity will improve the current IACTs one above a few TeV, and will be competitive with respect to the CTAO North one during the science verification phase above a few tens of TeV.

We note that the ASTRI Mini-Array performance remains almost stable across the 10\textdegree{} FoV, with a degradation at its edge by roughly a factor of two~\citep{2022JHEAp..35....1V,2022icrc.confE.884L}. On the other side, the EASs have 2\,sr field of view, but their energy and angular resolution in the same energy range as the ASTRI Mini-Array (about 10\,TeV) are at least a factor of 3--4 worse, making the ASTRI Mini-Array extremely competitive in studying the morphology of extended sources, crowded fields, and accurately monitoring multiple targets in the same pointing.

%
%
\section{The ASTRI Mini-Array Scientific Pillars}
%
%
The ASTRI Mini-Array science program can be divided into two periods. During the first four years of operations, the ASTRI Mini-Array will be run as an {\it experiment}, while in the second four-year period it will gradually evolve into an {\it observatory} open to the scientific community. 
%
%
\begin{figure*}[!ht]
	\centering
	\includegraphics[width=0.65\textwidth, angle=0]{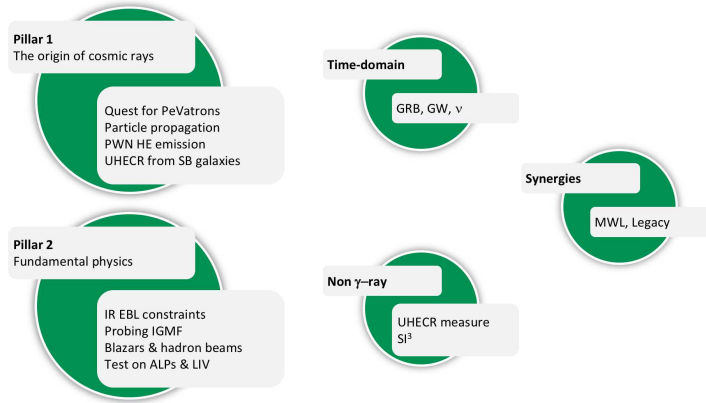}
	\caption{Graphical description of the ASTRI Mini-Array Core Science program.}
		\label{FIG:PILLARS}
\end{figure*}

Figure~\ref{FIG:PILLARS} shows a graphical description of the main Core Science topics that we plan to investigate during the first four years of operations. Our Core Science Program is based on ``Main Pillars''. They are science fields in which the ASTRI Mini-Array will contribute breakthrough pieces of evidence to improve our understanding of a few key science questions. 

{\bf Pillar-1} is mostly devoted to Galactic astrophysics and in particular it will address the study of {\it the origin of cosmic rays}.  The LHAASO Collaboration~\cite{2021Natur.594...33C} reported the discovery of twelve Galactic sources emitting gamma-rays at several hundreds of TeV up to 1.4\,PeV. These sources are able to accelerate protons up to $\sim 10^{15}$\,eV, making them ``PeVatron candidates". We note that the majority of these sources are diffuse \gray{} structures with angular extensions up to 1\textdegree{}, which, together with the LHAASO limited angular resolution, make the identification of the actual sources responsible for the ultra high-energy \gray{} emission not univocal (except for the Crab Nebula). 
%
This discovery is extremely important for the ASTRI Mini-Array science, especially because of its angular resolution which, at energies of about 100\,TeV, is a factor 3--4 times better in radius than the LHAASO one: 0.08\textdegree{} vs. 0.24\textdegree{} -- 0.32\textdegree{}. The ASTRI Mini-Array will investigate these and future PeVatron sources providing important information on their morphology 
above 10\,TeV.
A detailed discussion of the ASTRI Mini-Array investigation of Galactic PeVatrons is presented in~\cite{CardilloThisProc}.


The ASTRI Mini-Array wide FoV will be extremely important in the investigation of extended regions and/or point-like sources. A single pointing will allow us to investigate the Galactic Center or the Cygnus regions. On these regions we can accumulate several hundreds of hours, by including also the period of moderate Moon contamination. This is crucial to investigate both the (energy dependent) morphology of the sources in these regions, and their possible variability on long time-scale. 
The high-energy boundaries of the ASTRI Mini-Array will also be important to study the Crab Nebula, the only Galactic PeVatron currently known~\citep{2021ChPhC..45h5002A,2021Science373..425C}. 
The origin of the Crab Nebula gamma-ray emission detected by LHAASO does not require a hadronic contribution, but cannot exclude it either. A deep ASTRI Mini-Array observation lasting about 500\,hr in the 100-300 TeV range should definitely be able to provide constraints on the proton component in this source. 

{\bf Pillar-2} investigates {\it cosmology and fundamental physics} by studying blazars, a subclass of active galactic nuclei with a relativistic jet pointing towards the observer~\citep{1979ApJ...232...34B}. Their fast flux variability can probe the physical processes of their innermost regions, especially during bright flares as reported by~\citep{2007ApJ...664L..71A} for PKS~2155$-$304. Nevertheless, investigating blazars during their steady or high emission states can provide extremely valuable information too. The extra-galactic background light (EBL) affects the \gray{} spectrum of sources at high (z \textgreater{} 0.1) redshift emitting in the E \textgreater{} 0.1\,TeV energy band, and it is a key quantity to understand the intrinsic spectrum ($F_{\rm obs}(E_{\gamma}) = F_{\rm int}(E_{\gamma}) \times e^{-\tau(E_{\gamma,z})}$, where $\tau(E_{\gamma,z})$ is the optical depth at energy $E_{\gamma}$ and redshift $z$, $F_{\rm obs}(E_{\gamma})$ and $F_{\rm int}(E_{\gamma})$ are the observed and intrinsic spectrum, respectively) emitted by a distant source. From the mid-IR to the far-IR, where the IR background intensity is maximal, EBL direct measurements are prevented by the overwhelming dominance of the local emission from both the Galaxy and our Solar system. The relation between the energy of the \gray{}, $E_{\rm TeV}$, and the wavelength of the target EBL photon, $\lambda_{\rm max}$ ($\lambda_{\rm max} \simeq 1.24 \times E_{\rm TeV}$\,[$\mu$m]) yields the possibility to probe the EBL in the $\sim$(10--30)\,$\mu$m regime, otherwise unaccessible, by means of the study of blazars spectra above a few tens of TeV. Promising candidates to constrain the EBL up to $\sim$100\,$\mu$m are low-redshift radio galaxies (e.g., M~87, IC~310, Centaurus~A) and local star-bursting and active galaxies (M~82, NGC~253, NGC~1068). An ASTRI Mini-Array observation of IC~310 ($z = 0.0189$) can allow us to detect such source up to a few tens of TeV.

Dedicated observations of the ASTRI Mini-Array at energies above few TeV will allow us to put the ``hadron beam'' (HB) model for extreme BL Lacs to the test, look for anomalies in the spectrum related to photon-ALP (axion-like particle) mixing, perform tests of the Lorentz invariance violation (LIV), and probe the inter-galactic magnetic fields (IGMFs). All these studies can be performed through relatively long pointing of few carefully selected targets. Deep ($\sim 200$\,hr) observations of the extreme BL Lacertae object 1ES~0229$+$200~\citep[][$z=0.14$]{2007A&A...475L...9A} will provide spectra which can help disentangling between the standard EBL absorption scenario and the HB one. The same target can be effectively exploited to investigate the IGMF in the interval $10^{-14.5} \le B/[\rm G] \le 10^{-12}$ for a coherence length of about 1\,Mpc.

{\bf Time-domain and multi-messenger astrophysics} is currently in its golden age. There were a few major breakthroughs in the recent years. Currently, at least three extra-galactic sources seem to show associations with neutrino emission, namely NGC~1068, PKS~1424$+$240 and TXS~0506$+$056~\citep[see][and references therein]{2022Sci...378..538I}, associated to different AGN classes: TeV emitting BL~Lac objects and one Seyfert galaxy. There are important constraints imposed from these detections on the presence and/or absence of \gray{} VHE emission~\cite[see, e.g.,][for NGC~1068]{2016A&A...596A..68L,2019APh...112...16L} which could be investigated by means of the ASTRI Mini-Array. GRBs have been confirmed as emitters at TeV energies by MAGIC~\cite{MAGIC2019a,MAGIC2019b} and H.E.S.S~\cite{2019Natur.575..464A}. Recently, LHAASO detected emission above 10\,TeV from GRB~221009A~\citep{2022GCN.32677....1H}. This detection challenges the standard physics model, because of the very strong \gray{} attenuation expected from the EBL for a source at $z \simeq 0.15$ at $E>$\,a few TeV, opening a new window for alternative hypotheses~\citep[see e.g.,][and references therein]{2022arXiv221106935G}. The energy range above a few TeVs is the {\it right spot} for the ASTRI Mini-Array. In~\cite{2022JHEAp..35....1V} we simulated a GRB~190114C-like source
with a prompt energy of $3 \times 10^{53}$\,erg 
placed at  $z=0.1$ and $z=0.25$ (GRB~221009A is at $z=0.15$) and we predicted the capability to detect photons up to and above 10\,TeV within few minutes from the burst.

{\bf Non-gamma-ray astrophysics.} The ASTRI Mini Array will allow the {\it direct measurement of cosmic rays}, by investigating cosmic ray heavy nuclei. The techniques (e.g., for heavy nuclei such as iron) relies on the identification of a single high intensity pixel in the camera images of the detected Extensive Air Shower. This pixel lies between the reconstructed shower direction and the center of gravity of the shower. Furthermore, for the first time, we are in a position to image bright stars in the visible light waveband at very high angular resolution using a technique known as {\it stellar intensity interferometry}, which will allow us, by means of dedicated instrumentation, to obtain sub-milliarcsecond images of the brightest nearby stars and their environments.
%
%
\section{Conclusions}
%
Starting from 2025, the ASTRI Mini-Array will start collecting scientific data. They will constitute a {\it legacy} for both current and future VHE facilities and other multi-wavelength observatories, in terms of light-curves, spectra, and high resolution images of extended sources. 
%
%
%
\acknowledgments 
This work was conducted in the context of the ASTRI Project. We gratefully acknowledge support from the people, agencies, and organisations listed here: \href{http://www.astri.inaf.it/en/library/}{http://www.astri.inaf.it/en/library/}. This paper went through the internal ASTRI review process. 
%
%
%
\bibliographystyle{JHEP}
\scriptsize
\providecommand{\href}[2]{#2}\begingroup\raggedright\endgroup

%

\end{document}